\documentclass[11pt]{article}
\usepackage[margin=1in]{geometry}
\usepackage{graphicx}
\usepackage[pdftex,colorlinks=true,linkcolor=blue,citecolor=blue,urlcolor=black]{hyperref}
\usepackage{amsmath, amsthm, amssymb,amsfonts,bbm}
\usepackage{subfigure}
\usepackage{comment}
\usepackage{soul}
\usepackage{url}
\usepackage{pdflscape}
\usepackage[ruled,lined,linesnumbered]{algorithm2e}
\usepackage{hyperref}

\newcommand{\bR}{\mathbb{R}}

\newcommand{\bZ}{\mathbb{Z}}

\newcommand{\bF}{\mathbb{F}}

\newcommand{\cA}{\mathcal{A}}

\newcommand{\cB}{\mathcal{B}}
\newcommand{\cH}{\mathcal{H}}

\newcommand{\cR}{\mathcal{R}}
\newcommand{\cG}{\mathcal{G}}

\newcommand{\1}{\mathbbm{1}}

\newcommand{\tr}[1]{{\rm tr}\left({#1}\right)}
\newcommand{\Tr}{{\rm tr}}

\DeclareMathOperator{\Cov}{Cov}
\DeclareMathOperator{\dist}{dist}

\newcommand{\half}{\frac{1}{2}}

\newcommand{\be}{\begin{equation}}
\newcommand{\ee}{\end{equation}}
\newcommand{\bea}{\begin{eqnarray}}
\newcommand{\eea}{\end{eqnarray}}
\newcommand{\bes}{\begin{equation*}}
\newcommand{\ees}{\end{equation*}}
\newcommand{\beas}{\begin{eqnarray*}}
\newcommand{\eeas}{\end{eqnarray*}}

\makeatletter
\newtheorem*{rep@theorem}{\rep@title}
\newcommand{\newreptheorem}[2]{%
\newenvironment{rep#1}[1]{%
 \def\rep@title{#2 \ref{##1} (restated)}%
 \begin{rep@theorem}}%
 {\end{rep@theorem}}}
\makeatother

\newtheorem{thm}{Theorem}
\newtheorem*{thm*}{Theorem}

\newtheorem*{lem*}{Lemma}

\newtheorem{definition}[thm]{Definition}

\newreptheorem{thm}{Theorem}
\newreptheorem{lem}{Lemma}

\newcommand{\mjk}[1]{{\color{blue}#1}}

\begin{document}

\title{Finite correlation length implies efficient preparation of quantum thermal states}
\author{\small Fernando G.S.L. Brand\~ao$^1$ and Michael J. Kastoryano$^2$\\[8pt]
{\small $^1$ IQIM, California Institute of Technology, Pasadena CA 91125, USA}\\
{\small $^2$ NBIA, Niels Bohr Institute, University of Copenhagen, 2100 Copenhagen, DK}
}
\date{\today}

\maketitle

\begin{abstract}
Preparing quantum thermal states on a quantum computer is in general a difficult task. We provide a procedure to prepare a thermal state on a quantum computer with a logarithmic depth circuit of local quantum channels assuming that the thermal state correlations satisfy the following two properties: (i) the correlations between two regions are exponentially decaying in the distance between the regions, and (ii) the thermal state is an approximate Markov state for shielded regions. We require both properties to hold for the thermal state of the Hamiltonian on any induced subgraph of the original lattice. Assumption (ii) is satisfied for all commuting Gibbs states, while assumption (i) is satisfied for every model above a critical temperature. Both assumptions are satisfied in one spatial dimension. Moreover, both assumptions are expected to hold above the thermal phase transition for models without any topological order at finite temperature.  As a building block, we show that exponential decay of correlation (for thermal states of Hamiltonians on all induced subgraphs) is sufficient to efficiently estimate the expectation value of a local observable. Our proof uses quantum belief propagation, a recent strengthening of strong sub-additivity, and naturally breaks down for states with topological order. 
 \end{abstract}

\section{Introduction}

Preparing, or sampling from, thermal (Gibbs) states is one of the most important tasks for the simulation of physical systems. The Metropolis sampling algorithm and its variants is the workhorse for classical statistical mechanical models. Unfortunately, no single purpose algorithm is known for sampling quantum thermal states, since transitions between energy levels of the system Hamiltonian cannot in general be performed by local operations. A number of quantum Gibbs sampling algorithms have been proposed \cite{GibbsSampling,qMetropolis,terhal2000,poulin2009}, with two algorithms standing out:  the quantum Metropolis algorithm \cite{qMetropolis}, and the Davies Gibbs sampling algorithm \cite{GibbsSampling}. The quantum Metropolis algorithm emulates the classical metropolis algorithm at the quantum circuit level. It is general purpose, and as such is the leading practical candidate for use on a programmable quantum computer, but its performance is extremely difficult to analyze. Furthermore, whereas the classical Metropolis algorithm closely models the action of a thermal reservoir on a physical system, the quantum Metropolis algorithm seems disconnected from a physical preparation mechanism. The Davies Gibbs sampling algorithm, on the other hand, models the physical process of a system weakly interacting with a thermal bath, and under certain circumstances it is possible to provide guarantees on its runtime \cite{GibbsSampling}. Unfortunately, the Davies Gibbs sampling algorithm only works for quantum Hamiltonians with locally commuting terms, which is a severe drawback \footnote{Davies Gibbs sampling can be defined for non-commuting models as well, but the Liouvillian is non-local in general.}.  

In this paper, we introduce a new Gibbs state preparation procedure for which we have efficiency guarantees that reflect physically reasonable assumptions. The simulation algorithm takes the form of a circuit of local quantum channels that approximates the desired Gibbs state exponentially well in the depth of the circuit.  It is based on the idea that if the Gibbs state has exponentially decaying correlations, then expectation values of observables in the bulk cannot be sensitive to fluctuations at the boundary. Therefore, patching bits of the thermal state together will ``look" like the global Gibbs state everywhere except where the patches were stitched together. If we then patch bits of the thermal state over the stitch lines, this will smoothen them out. Such a procedure works under two assumptions: that correlations are exponentially decaying, and  that erased patches can be recovered locally. 

Our preparation scheme builds upon recent progress in the fields of quantum information and many-body theory. In particular, we use the formalism and intuition that was developed in Ref. \cite{GibbsSampling} for analyzing quantum Gibbs states. This was in turn inspired by a large body of work analyzing the convergence behavior of Glauber dynamics for classical spin systems \cite{ziggy,martinelli,diaconis}. It is interesting to note that our proof applied to a classical spin system is novel, and could provide new insight in their analysis. We also make use of the recent strengthening of strong subaditivity, where it was shown that if the conditional mutual information between three disconnected regions $ABC$ is small,  there exists a quantum channel recovering the erasure of region $A$ from the information in $B$ only \cite{FR2015,junge2015}. Finally, we need the quantum belief propagation equations derived in Ref. \cite{hastingsBP}. All of these ingredients when put together make our proof relatively simple and very intuitive, contrary to some other convergence proofs based on a detailed analysis of the spectral gap of a many-body generator. 

Finally, our algorithm is formulated and proved to converge rapidly for certain classes of Gibbs states, but it also works as a pure state preparation algorithm if the pure state is the ground state of a local frustration-free Hamiltonian. Crucially, the assumptions in the theorem break down whenever the pure state has a non-zero topological entanglement entropy. This is necessary, since the algorithm converges in a number of local steps thats scales as the logarithm of the system size, whereas we know that no topologically ordered state can be prepared with a finite depth circuit of quantum channels \cite{konig2014}. 
The scaling of our algorithm is similar to that for the preparation of injective PEPS with a uniformly gapped parent Hamiltonian in Ref. \cite{ge2016}. We believe that there are connections between the two algorithms, that otherwise appear to be very distinct, although we do not elaborate on them in this paper.

\vspace{0.2 cm}

\noindent \textbf{Summary of results}: 
We start by introducing the necessary framework to formulate the two main assumptions in the paper: (i) uniform clustering of correlations, and (ii) the uniform Markov property. For a given lattice, and a Gibbs state defined on it, the system is {\it uniformly clustering} if for any subset of the lattice, any two observables defined on the subset have exponentially decaying correlation as defined by the covariance (see Def. \ref{def:clustering}). The definition is distinct from the more conventional notion as we require that the Gibbs state defined on a reduced region (rather than the reduction of the total Gibbs state to the region) satisfies some form of decay of correlations. A system is said to satisfy the uniform Markov property if for any region $A$, the conditional mutual information between $A$ and $C$, where $B$ shields $A$ from $C$, is decaying in the width of $B$. 

The first main result is that expectation values of local observables can be estimated locally if the state is uniformly clustering. The ability to locally estimate expectation values is often referred to as local indistinguishability, and is closely related to the notion of local topological quantum order for ground states, as defined in Ref. \cite{thespiros}. Building on the first result, we prove the main result of the paper: If the Gibbs state is (i) uniformly clustering, and (ii) uniformly Markovian for non-contractible regions of the lattice, then we can construct a constant depth circuit of quasi-local quantum channels that prepares the Gibbs state from any initial state. The precision of the approximation improves exponentially with the support of the quasi-local circuit elements, so we can choose the locality of the circuit to be $O(\log(N))$, where $N$ is the number of particles, if the error in the uniform clustering and Markov conditions is exponentially decaying (see Theorem \ref{thm:main}). As a corollary of our main theorem, we show that under the same assumptions  the procedure can be iterated to yield a log-depth circuit of strictly local quantum channels to prepare the Gibbs state.   

We point out the following extensions or sharpening of our main theorem: (a) in one dimension, both assumptions are satisfied and hence all one dimensional Gibbs states of local bounded Hamiltonians can be prepared efficiently on a quantum computer. (b) We find that  the Gibbs state of commuting Hamiltonians always satisfy the uniform Markov property, and hence the efficiency of preparation only depends on uniform clustering. This also implies that quantum memories based on commuting projector codes, such as the 4D Toric code, must have long range correlations. (c) Finally, we note that our main proof can be extended to the preparation of ground states of gapped frustration-free Hamiltonians. In the case of stabilizer Hamiltonians, the ability to efficiently prepare the state is guaranteed by a trivial topological entanglement entropy; hence explicitly connecting the circuit definition of topological order and the topological entanglement definition\footnote{A similar construction was discovered earlier by Alexei Kitaev and presented in the lecture \href{http://scgp.stonybrook.edu/video_portal/video.php?id=2010}{\mjk{available online}}.}.  

\section{Preliminaries}

\subsection{Notation}
This paper concerns quantum spin  systems on the lattice. Although the results presented here can be extended to more general graphs, we will restrict our attention to spins living on a $D$-dimensional finite square lattice $\Lambda \subseteq \bZ^D$, which can be identified with $(\bZ \backslash L)^D$ for an integer $L$ (we will call $L$ the lattice side length). Lattice subsets will be denoted by upper case Latin letters, e.g. $A,B\subseteq\Lambda$. The complement of a set $A\in\Lambda$ will be written $A^c$. The cardinality of a set $A$ will be denoted by $|A|$. 
The  Hilbert space associated to a subset  $A\subseteq\Lambda$ is $\cH_A=\bigotimes_{x\in A}\cH_x$. We assume the local Hilbert spaces are finite dimensional, i.e. ${\rm dim}(\cH_x)<\infty$. For any $A\subseteq\Lambda$, we denote the set of bounded operators acting on $\cH_A$ by $\cB_A$, and its Hermitian subset by $\cA_A\subseteq\cB_A$. The elements of $\cA_\Lambda$ will be called observables, and will always be denoted with lower case Latin letters ($f,g\in\cA_\Lambda$). 
We will say that an operator $f \in \cB_\Lambda$ has support on $A \subset \Lambda$ if it can be written as $f_{A} \otimes \1_{A^c}$, for an operator $f_A \in \cB_A$. For $i,j\in\Lambda$, we denote by $\dist(i,j)$ the Euclidean distance in $\bZ^d$. The distance between two sets $A,B\subseteq\Lambda$ is $\dist(A,B) :=\min\{\dist(i,j),i\in A,j\in B\}$. We will write the operator norm of $f$ as $||f||$, the trace norm is written $||\cdot||_1$. 

The main object of study in this paper is the thermal (or Gibbs) state of some global Hamiltonian $H\in\cA_\Lambda$. The Hamiltonian can be written as a sum of hermitian elements $h^Z\in\cA_\Lambda$ with support on the subset $Z\subseteq\Lambda$ as $H=\sum_{Z\subset\Lambda} h^Z$. The Hamiltonian is called \textit{local} if for every $Z$ with a diameter larger than a given constant $r$ we have $h^Z=0$. A local Hamiltonian  is \textit{bounded} if there exists a constant $K$ such that for each of its local terms, $||h^Z||\leq K$. 
For any $A\subset\Lambda$, we define $H^A=\sum_{Z\subset A}h^Z$ to be all of the terms of $H$ with support within $A$. The terms of the Hamiltonian intersecting the boundary of $A$ are labelled $H^{\partial A}=\sum_{Z s.t. Z\subset A \wedge Z\subset A^c}h^Z$.  The thermal (Gibbs) state of the full lattice $\Lambda$ is

\be \rho\equiv\rho^\Lambda= e^{-\beta H^\Lambda}/\tr{e^{-\beta H^\Lambda}},\ee
where $\beta\in \bR^+$ is the inverse temperature. 
Restricted Gibbs states will similarly be written 
$\rho^A=e^{-\beta H^A}/\tr{e^{-\beta H^A}}$,
for any $A\subseteq\Lambda$. We will always write a superscript to mean the region of the lattice on which the Gibbs state is defined. A subscript on states refers to the partial trace as usual; for example $\rho^{AA^c}_A=\Tr_{A^c}[\rho^{AA^c}]$. Unless otherwise specified, $\rho$ will always refer to the Gibbs state. The covariance of two observables $f,g$ in a state $\sigma$ is given by:
\be \Cov_\sigma(f,g)=|\tr{\sigma f g}-\tr{\sigma f}\tr{\sigma g}|.\ee

\subsection{Clustering}

The main assumptions throughout the paper will be on the structure of the correlations in the thermal state. We define below an extension of the usual two point function correlation decay: 
\begin{definition}[Uniform clustering]\label{def:clustering}
Let $H$ be a local bounded Hamiltonian,  and $\rho$ the Gibbs state of $H$ at inverse temperature $\beta>0$. The pair $(H, \beta)$ is said to be {\rm uniformly $\epsilon(\ell)$-clustering} if for any $X\subset\Lambda$ and any $A\subset X$ and $B\subset X$ such that $\dist(A,B)\geq \ell$, 
\be\Cov_{\rho^X}(f,g)\leq ||f||~||g|| \epsilon(\ell)\label{eqn:clustering}\ee
for any $f$ supported on $A$ and $g$ supported on $B$. 
\end{definition}

$\epsilon(\ell)$ will typically be a monotonically decreasing function of the distance $\ell$. 
If a specific state $\rho^X$ satisfies Eqn. (\ref{eqn:clustering}), then we simply say that $\rho^X$ is $\epsilon(\ell)$-clustering.  We call this property uniform clustering to contrast with regular clustering property \cite{GibbsSampling} that usually only refers to properties of the bulk state $\rho^\Lambda$. To the best of our knowledge, the uniform clustering assumption has not been considered in the classical setting. It is as far as we can tell strictly different from the notions of 'strong' and 'weak' mixing in the classical Gibbs sampler setting. There one wants to distinguish between the situations where correlations decay from the source of a perturbation of the boundary or from the nearest boundary point \cite{martinelli19}. This does not appear to be relevant to the notion of uniform clustering above, since here we compare restricted lattices with 'open boundary conditions'. 

We would expect that regular clustering implies uniform clustering for classical spin lattice models, although we were not able to show this. In the quantum setting, on the other hand, there is clearly a difference. 
For topologically ordered systems, such as the Toric \cite{kitaev2003} or Color \cite{bombin2006} code Hamiltonians, the ground states are  clustering for {\it contractible} regions, however such systems are not  clustering if we allow for {\it non-contractible} regions. Hence, these provide examples of quantum systems the are globally clustering but not uniformly clustering. In the proof of Theorem \ref{thm:main} we will crucially need the Gibbs state to be uniformly clustering also for non-contractible regions. Classical Gibbs states do not make a distinction between contractible and non-contractible regions, as they cannot exhibit topological order. Whether there is a distinction for Gibbs states of quantum systems is still an open question.  We discuss this point in more detail in Sec. \ref{sec:implications}.

\subsection{Quantum Belief Propagation}
If the Hamiltonian $H=H_0+V$ satisfies $[H_0,V]=0$, then the Gibbs state of $H$ can be expressed in terms of the Gibbs state of $H_0$ as $e^{-\beta H}=e^{-\beta H_0}e^{-\beta V}$. The same is not true if $H_0$ and $V$ do not commute. However, if the hamiltonian is a sum of local terms, this is approximately true if the perturbation is also local \cite{hastingsBP,kim2012perturbative,KatoBrandao16}. 

Let $H_0$ be a local bounded Hamiltonian, and let $V$ be a bounded Hermitian operator with finite local support. Define the perturbed Hamiltonian $H(s)=H_0 +sV$, with $0\leq s \leq 1$. In Ref. \cite{hastingsBP}, it was shown that 
\be \frac{\partial}{\partial s}e^{-\beta H(s)}= \xi_s e^{-\beta H(s)}+e^{-\beta H(s)}(\xi_s)^\dag, \ee
where $\xi_s=-\frac{\beta}{2}\int d\omega (1+G(\omega))$ and $G(\omega)=(\beta \omega/2)^{-1}-\coth(\beta\omega/2)$. In a slight modification of the proof in \cite{hastingsBP} (see \cite{KatoBrandao16} Sec. IIIa), one easily obtains
\be \frac{\partial}{\partial s}e^{-\beta H(s)}= \zeta_s e^{-\beta H(s)}+e^{-\beta H(s)}(\zeta_s)^\dag, \label{eqn:BP2}\ee
where $\zeta_s=\frac{\beta}{2}\int d\omega\frac{\tanh(\beta\omega/2)}{\beta\omega/2} V^{\omega,s}$, and $(V^{\omega,s})_{ab}=V(s)_{ab}(E_a(s)-E_b(s)-\omega)$, with $E_a(s),E_b(\omega)$ eigenvalues of $H(s)$ and $V(s)_{ab}$ matrix elements of $V$ in the eigenbasis of $H(s)$.  Solving Eqn. (\ref{eqn:BP2}), yields 
\be e^{-\beta H(s)}=\eta^V(s)e^{-\beta H_0}\eta^V(s)^{\dag},\ee
where $\eta^V(s)=\mathcal{T}e^{\frac{\beta}{2}\int_0^s d\lambda \zeta_\lambda}$ and $\mathcal{T}e^{(\cdots)}$ denotes the time-ordered exponential. 

In Ref. \cite{KatoBrandao16} (Sec. IIIa) it was shown that 
\be ||\eta^V(s)||\leq e^{\frac{\beta}{2}||V||},\ee for all $0\leq s\leq 1$, and that there exists an operator $\eta_\ell^V(s)$ with support on a ball of radius $\ell$ around the support of $V$ such that 
\be||\eta^V(s)-\eta_\ell^V(s)||\leq e^{c_1||V||}e^{-c_2\ell},\ee 
for some constants $c_1,c_2$ depend upon the inverse temperature and on the Lieb-Robinson velocity of the model, but not on the volume $|\Lambda|$ (the constants are written out explicitly in Ref. \cite{KatoBrandao16}). Writing $\eta^V\equiv \eta^V(1)$ and $\eta_\ell^V\equiv \eta_\ell^V(1)$, we can summarize these results in the following proposition:

\begin{thm}[Quantum Belief Propagation \cite{hastingsBP}]\label{thm:BP}
Let $H_0$ be a local bounded Hamiltonian,  and let $V$ be a bounded Hermitian operator with finite local support. For $H=H_0+V$, there exists an operator $\eta^V$, satisfying $||\eta^V||\leq e^{\beta/2||V||}$ and $e^{-\beta H}=\eta^Ve^{-\beta H_0}\eta^{V,\dag}$ such that 
\be ||\eta^V-\eta_\ell^{V}||\leq e^{c_1||V||}e^{-c_2\ell},\label{eqn:bf1},\ee 
for some positive  constant $c_1, c_2$, where $\eta^{V}_\ell$ is an operator with support on a ball of radius $\ell$ around the support of $V$. 
\end{thm}
An important corollary of Theorem \ref{thm:BP} is that for a bounded local perturbation $V$, we get 

\be ||e^{-\beta (H_0+V)}-\eta^V_\ell e^{-\beta H_0}\eta_\ell^{V,\dag}||_1\leq C_1 e^{-c_2\ell}:=\gamma(\ell),\label{eqn:beliefprop},\ee
for some constants $C_1,c_2$. 
 The bound is always exponential for local $H$ and a local perturbation $V$. We will write it as $\gamma(\ell)$ for later convenience to keep track of all error terms explicitly. 
A derivation of the belief propagation equations can be found in Refs. \cite{hastingsBP,KatoBrandao16,kim2012perturbative} The actual form of the $\eta_\ell^V$ is not important to us now, rather we only need their locality and boundedness properties. The quantum belief propagation equations have proved to be a very useful tool in the numerical analysis of quantum Gibbs states \cite{hastingsBP,poulin2008,bilgin2010}. They have also been used to show the efficient preparation of Gibbs states in one dimension in combination with a block decimation technique \cite{bilgin2010prep}.

\subsection{The Markov property and recovery maps}

\begin{figure}
\centering
  \includegraphics[scale=0.40]{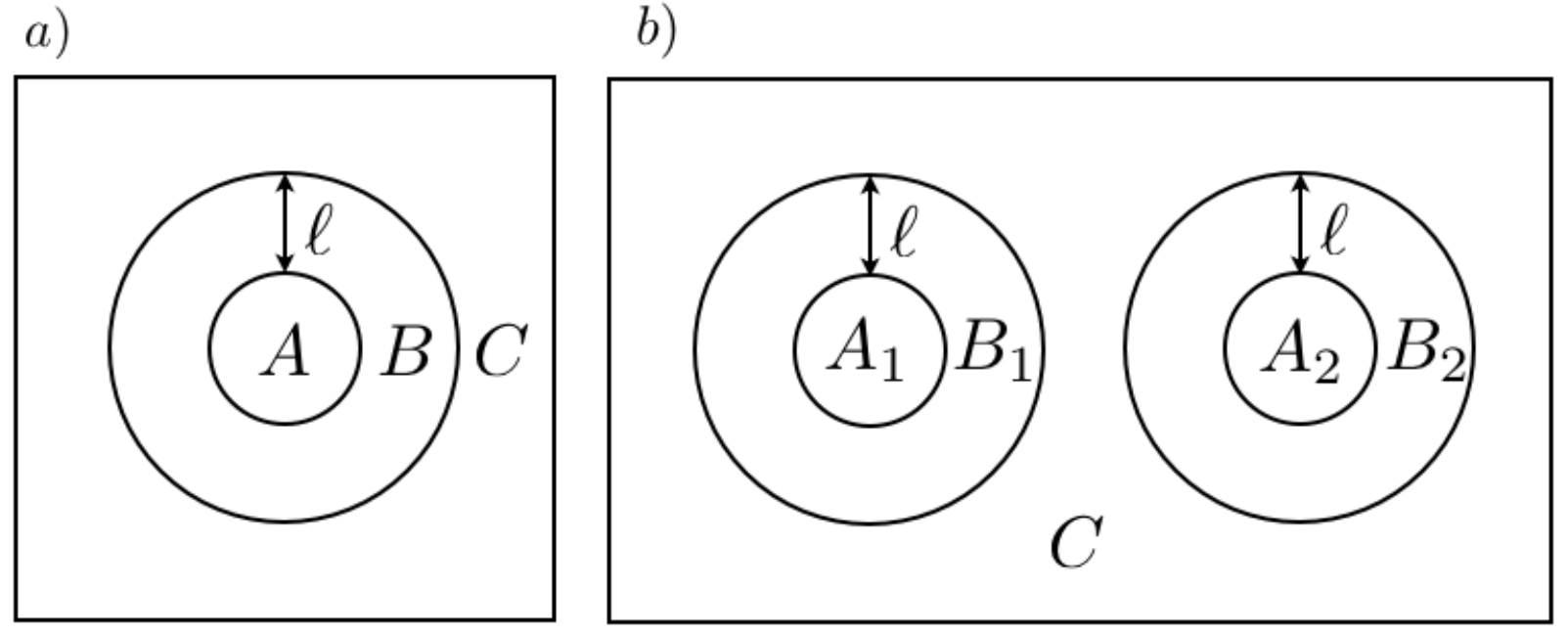}
    \caption{Two different lattice decompositions illustrating the setup for the uniform Markov property.  a) System $ABC=\Lambda$ consists of a circle $A$ shielded from $C$ by a ring $B$ of width $\ell$. b) System $A_1B_1A_2B_2C$ consists of two circles each shielded from $C$ by a ring of width $\ell$. The outer circles do not overlap. The union property says that if regions $A_1,A_2$ can be recovered locally, and their recovery operations do not overlap, then the union of the regions can be recovered locally.  }
    \label{fig}
\end{figure}

Gibbs states of local Hamiltonian satisfy an \textit{Area Law} for mutual information \cite{wolf2008area}, meaning that for any $A\subset\Lambda$,
\be I_\rho(A:A^c)= O( |\partial A|),\label{eqn:area}\ee
where $I_\rho(A:B)=S_\rho(A)+S_\rho(B)-S_\rho(AB)$ is the mutual information of the Gibbs state $\rho$ and $S_\rho(A)=-\tr{\rho_A\log(\rho_A)}$ is the Von Neumann entropy of the reduced state $\rho_A$. 

The conditional mutual information is a measure of correlations between two quantum systems from the perspective of a third one. It is defined as
\be I_\rho(A:C|B)=S_\rho(AB)+S_\rho(BC)-S_\rho(B)-S_\rho(ABC).\ee
The conditional mutual information is always positive because of the strong subadditivity of Von Neumann entropy. 
On top of guaranteeing that an area law holds, a small conditional mutual information actually bounds the rate at which the mutual information saturates with a growing $B$ region (see Ref. \cite{KatoBrandao16}). We will want to characterize systems with a small conditional mutual information.

One of the main implications of the conditional mutual information being small is that it implies the existence of a local recovery map:

\begin{thm}[Fawzi-Renner \cite{FR2015}] \label{thm:FR}
For any state $\sigma$ on $ABC=\Lambda$ as in Fig. \ref{fig}a), there exists a quantum channel $\cR_B^{AB}: B \rightarrow AB$ such that 
\be I_\sigma(A:C|B)\geq -2\log_2 F(\sigma,\cR_B^{AB}(\sigma_{BC})),\label{eqn:FR}\ee
where $F(\rho,\sigma):=||\sqrt{\rho}\sqrt{\sigma}||_1$ is the fidelity, and $\sigma_{BC}=\Tr_A[\sigma]$.
\end{thm}

In terms of the trace norm, Eqn. (\ref{eqn:FR}) reads
\be I_\sigma(A:C|B)\geq\frac{1}{4\log{2}}||\sigma-\cR_B^{AB}(\sigma_{BC})||^2_1.\ee
In Ref. \cite{junge2015}, it was shown that the recovery map $\cR_B^{AB}$ can always be chosen as a rotated Petz map \cite{petz}. In the following, we will not need to invoke its explicit form. Rather its existance is sufficient for our uses. The converse of Theorem \ref{thm:FR} is also true by the Fannes-Audenaert inequality \cite{Fannes}:
\be I_\sigma(A:C|B)\leq 3 \log(d_{AB}) ||\sigma^{ABC}-\cR_{AB}(\sigma_{BC})||^{\half}_1,\ee
where $d_{AB}$ is the dimension of the Hilbert space associated to the lattice subsets $AB$. For a qubit lattice, $d_{AB}=2^{|AB|}$.

One immediate consequence of Theorem \ref{thm:FR} is that if two sufficiently separated regions are traced out, and each can individually be recovered, then both regions can be recovered. In analogy to quantum error correcting codes, we call this the \textit{union property} (see Fig. \ref{fig}b):

Consider regions $A_1A_2B_1B_2C=\Lambda$, such that $A_1$ and $A_2$ are separated by a distance at least $2\ell$, and $B_1,B_2$ shield $A_1,A_2$ from $C$ by a distance at least $\ell$ each. If there exist recovery maps $\cR_{A_1B_1}$ and $\cR_{A_2B_2}$ such that $||\sigma-\cR_{A_1B_1}(\sigma_{B_1A_2B_2C})||_1\leq \epsilon$ and $||\sigma-\cR_{A_2B_2}(\sigma_{B_2A_1B_1C})||_1\leq \epsilon$, then 
\be ||\sigma-\cR_{AB}(\sigma_{BC})||_1\leq 2 \epsilon,\ee
by the triangle inequality, and the fact that the recovery maps commute, and where  we write $A\equiv A_1A_2$ and $B\equiv B_1B_2$ for convenience. 
For the remainder of the paper, we will use the following formal definition:

\begin{definition}
[uniform Markov property]\label{def:markov} Let $H$ be a local bounded Hamiltonian, and $\rho$ the Gibbs state of $H$ at inverse temperature $\beta$. The pair $(H, \beta)$ is said to satisfy the {\rm uniform $\delta(\ell)$-Markov condition} if for any $ABC=X\subset\Lambda$, such that $\dist(i,j)\geq \ell$ for any $i\in A$ and $j\in C$, we have
\begin{equation}
I_{\rho^X}(A : C | B) \leq \delta(\ell).
\end{equation}
\end{definition}

\subsection{Approximation Errors}

There are three main approximation errors which all appear in the formulations of Theorems \ref{thm:local_indist} and Theorem \ref{thm:main} below: from the uniform Markov property $\delta(\ell)$, from belief propagation $\gamma(\ell)$ and from the uniform clustering condition $\epsilon(\ell)$. For a local perturbation, we know that $\gamma(\ell)\leq c_1e^{-c_2\ell}$ always holds for some constants $c_1,c_2$. The decay of $\epsilon(\ell)$ and $\delta(\ell)$ will be taken as assumptions. Typically for non-critical thermal many body systems,    $\epsilon(\ell)$ and $\delta(\ell)$ will be exponentially decaying in $\ell$, but we will only need them to be decaying faster then $1/{\rm poly}(\ell)$. 

If the Hamiltonian consists of locally commuting terms, as do stabilizer \cite{gottesman1997} or Levin-Wen \cite{levinwen} Hamiltonian, then beyond a fixed distance $\ell_{\rm c}$ determined by the maximal range of the Hamiltonian, the approximation errors $\gamma(\ell)=\delta(\ell)=0$ for all $\ell>\ell_c$. This means that for commuting Hamiltonians, the only approximation error comes from uniform clustering \footnote{Some special systems, like Graph state Hamiltonians \cite{hein2004} also have a critical distance beyond which $\epsilon(\ell)=0$, but these are somehow very special systems that don't reflect general properties of non-critical statistical mechanics models.}  

\section{Results}

\begin{figure}
\centering
  \includegraphics[scale=0.50]{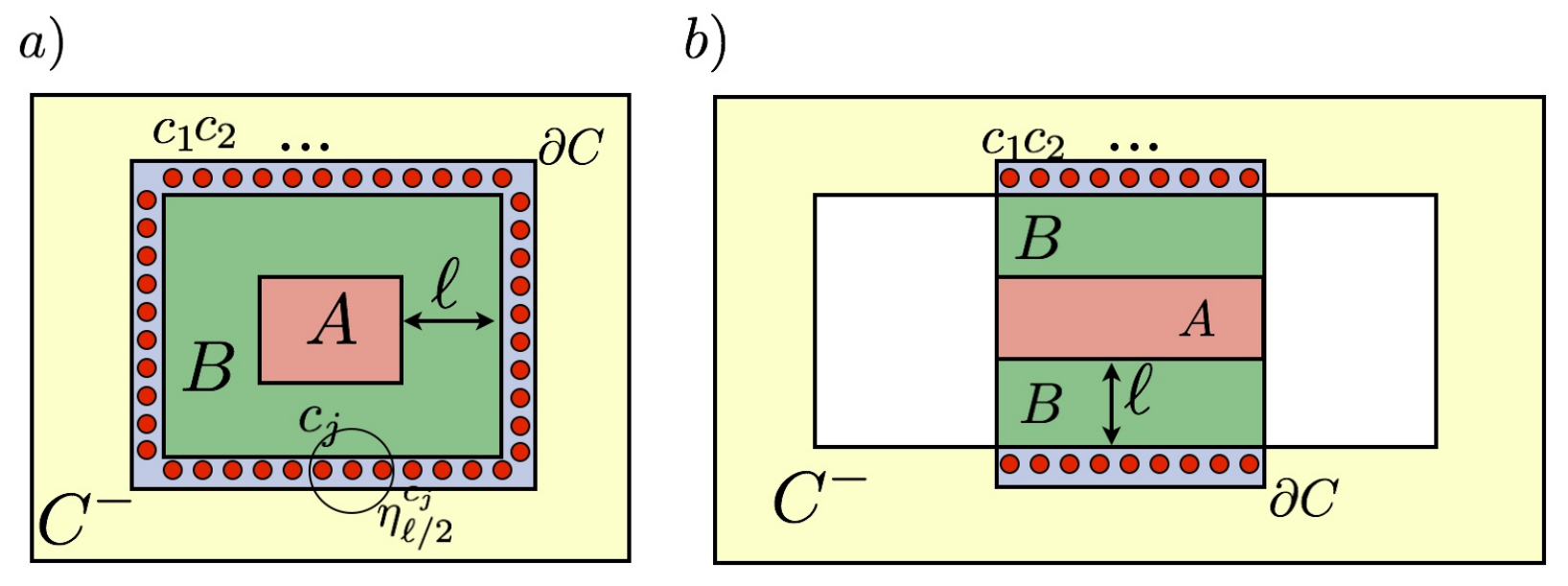}
    \caption{Two different lattice decompositions in terms of disjoint subsets $ABC=X\subset \Lambda$. The red circles represent the boundary vertices $\partial C$ that boarder with region $B$ (green). We write $C^-=C\setminus\partial C$ to mean the interior of $C$ (yellow). The terms of the hamiltonian that overlap with one particular circle can be removed with an operator that acts quasi-locally around it. The main idea of the proof of Theorem \ref{thm:local_indist} is to remove the terms intersection the boundary one by one. In a), the region $A$ (light red) is contractible, whereas in b) it is non-contractible. The difference is clear, as in a) we can remove all terms in a closed ring around $A$, whereas this is not possible in b) without including two holes, thus changing the topology of the problem.}
    \label{fig1}
\end{figure}

\subsection{Estimating local expectation values}

In this section, we prove the first main result of the paper, which says that if a system is uniformly clustering, then the expectation value of an observable on a contractible region can be estimated locally. More specifically:

\begin{thm}[Local indistinguishability]\label{thm:local_indist}
Let $H$ be a local bounded Hamiltonian,   and $\rho$ the Gibbs state of $H$ at inverse temperature $\beta$. Let $A B C=X\subset \Lambda$ be such that $\dist(i,j)\geq \ell$ for any $i\in A$ and $j\in C$ (see Fig. \ref{fig1}). Let $\rho^X$ be the Gibbs state of $H^X$, and let $\rho^{AB}$ be the Gibbs state on $AB$. If the system is uniformly $\epsilon(\ell)$-clustering, then 
\be||\Tr_{BC}(\rho^X)-\Tr_{B}(\rho^{AB})||_1\leq K |\partial C| (\epsilon(\ell/2)+\gamma(\ell/2)),\label{eqn:localindist}\ee
for some constant $K$, where $\partial C$ is the boundary of $C$ with $B$.
\end{thm}
\proof{
We start by labeling all of the sites of $C$ neighboring $B$ as $c_j\in\partial C$, with $j=1,...,M$ and $M=|\partial C|$. We will want to consider the sequence of Hamiltonians defined on the graphs $X_k=X\setminus\cup_{j=1}^k c_j$, where we remove elements of $\partial C$ one by one. In this notation, $X_0=X\equiv ABC$ and $X_M=X\setminus\partial C\equiv AB\cup C^-$, where $AB$ and $C^-=C\setminus\partial C$ are disjoint. See Fig. \ref{fig1} for an illustration. We will consider the sequence of Hamiltonians $H^{X_j}$ for $j=1,...,M$, and the sequence of Gibbs states $\rho^{X_j}$. Because of the uniform $\epsilon(\ell)$-clustering assumption, every $\rho^{X_j}$ is $\epsilon(\ell)$-clustering.

 Note that $\Tr_{B}(\rho^{AB})=\Tr_{BC}(\rho^{AB}\otimes\rho^{C^-})$, hence we can expand the trace norm difference as a telescopic sum
\bea ||\Tr_{BC}(\rho^X)-\Tr_{B}(\rho^{AB})||_1&=& ||\Tr_{BC}(\rho^X-\rho^{AB}\otimes\rho^{C^-})||_1\nonumber\\
&\leq& \sum_{j=0}^{M-1} ||\Tr_{BC}(\rho^{X_{j+1}}-\rho^{X_j})||_1\eea

Now, note that $\rho^{X_{j+1}}$ and $\rho^{X_j}$ differ by a constant number of local Hamiltonian term overlapping with $c_{j+1}$ on the inner boundary of $C$, hence from the belief propagation equations (\ref{eqn:beliefprop}) there exists an operator $\eta^{c_{j+1}}_{\ell/2}$ such that 
\be ||\rho^{X_{j+1}}-\eta_{\ell/2}^{c_{j+1}}\rho^{X_{j}}\eta_{\ell/2}^{c_{j+1},\dag}||_1\leq \gamma(\ell/2).\ee
$\eta^{c_{j+1}}_{\ell/2}$ has support on a ring of width $\ell/2$ around site $c_{j+1}$, and $\gamma(\ell)$ is exponentially decaying (see Eqn. (\ref{eqn:beliefprop}). 
We can bound each individual term as follows:

\bea ||\Tr_{BC}(\rho^{X_j}-\rho^{X_{j+1}})||_1&=&\sup_{||g_A||\leq 1} \left|\tr{g_A(\rho^{X_j}-\eta^{c_{j+1}}\rho^{X_j}\eta^{{c_{j+1}},\dag})}\right|\nonumber\\
&\leq&\sup_{||g_A||\leq 1} \left|\tr{g_A(\rho^{X_j}-\eta_{\ell/2}^{c_{j+1}}\rho^{X_j}\eta_{\ell/2}^{{c_{j+1}},\dag})}\right|+2\gamma(\ell/2)\nonumber\\
&=& \sup_{||g_A||\leq 1} \Cov_{\rho^{X_j}}(g_A, \eta_\ell^{c_{j+1},\dag}\eta_\ell^{c_{j+1}})+2\gamma(\ell/2)\nonumber\\
&\leq&  \epsilon(\ell/2)||\eta_\ell^{c_{j+1},\dag}\eta_\ell^{c_{j+1}}|| +2\gamma(\ell/2)\nonumber\\
&\leq& c(\epsilon(\ell/2)+\gamma(\ell/2)),\label{eqn:step1cl}\eea
since $c\equiv||\eta^{c_{j+1},\dag}_\ell\eta_\ell^{c_{j+1}}||\geq2$ is a constant. The $\ell/2$ argument in the functions $\epsilon$ and $\gamma$ is there because the support of $\eta_{\ell/2}^{c_{j+1}}$ is only a distance $\ell/2$ away from $A$. Summing all of the terms in the telescopic sum gives a pre-factor of order $|\partial C|$. \qed}

Eqn. (\ref{eqn:localindist}) is generally referred to as \textit{local indistinguishability}, and is closely related to the stability of gapped phases \cite{thespiros}. It tells us that as long as $\epsilon(\ell)$ is decaying sufficiently rapidly in $\ell$, then expectation values of local observables can be evaluated quasi-locally. Theorem \ref{thm:local_indist} does not impose any restrictions on the shape of $X$ or on the structure of $f$ and $g$. Rather it only relies on the uniform clustering of $\rho$. In the proof of Theorem \ref{thm:main}, we will need $\rho^X$ to be clustering for a topologically non-trivial subset $X$, and for non-contractible regions $ABC=X$. This is necessary for Theorem \ref{thm:main} to hold as discussed in Sec. \ref{sec:implications}. Finally, we note that local indistinguishability also satisfies a union property. This is simply inherited by the union property of covariances. A proof very similar to the one above  was recently used to show that expectation values of local observables for injective PEPS with a uniformly gapped parent Hamiltonian can be efficiently estimated \cite{schwarz2016}. 

The bound in Enq. (\ref{eqn:clustering}) can be strengthened in the case when the hamiltonian has locally commuting terms. Then $\gamma(\ell)$ is zero beyond some range of the Hamiltonian. Hence, we do not need to use the first step in Eqn. \ref{eqn:step1cl}, and we are left with an error term scaling as $\epsilon(\ell)$ rather than $\epsilon(\ell/2)$.

\subsection{The main theorem}

\begin{figure}
\centering
  \includegraphics[scale=0.40]{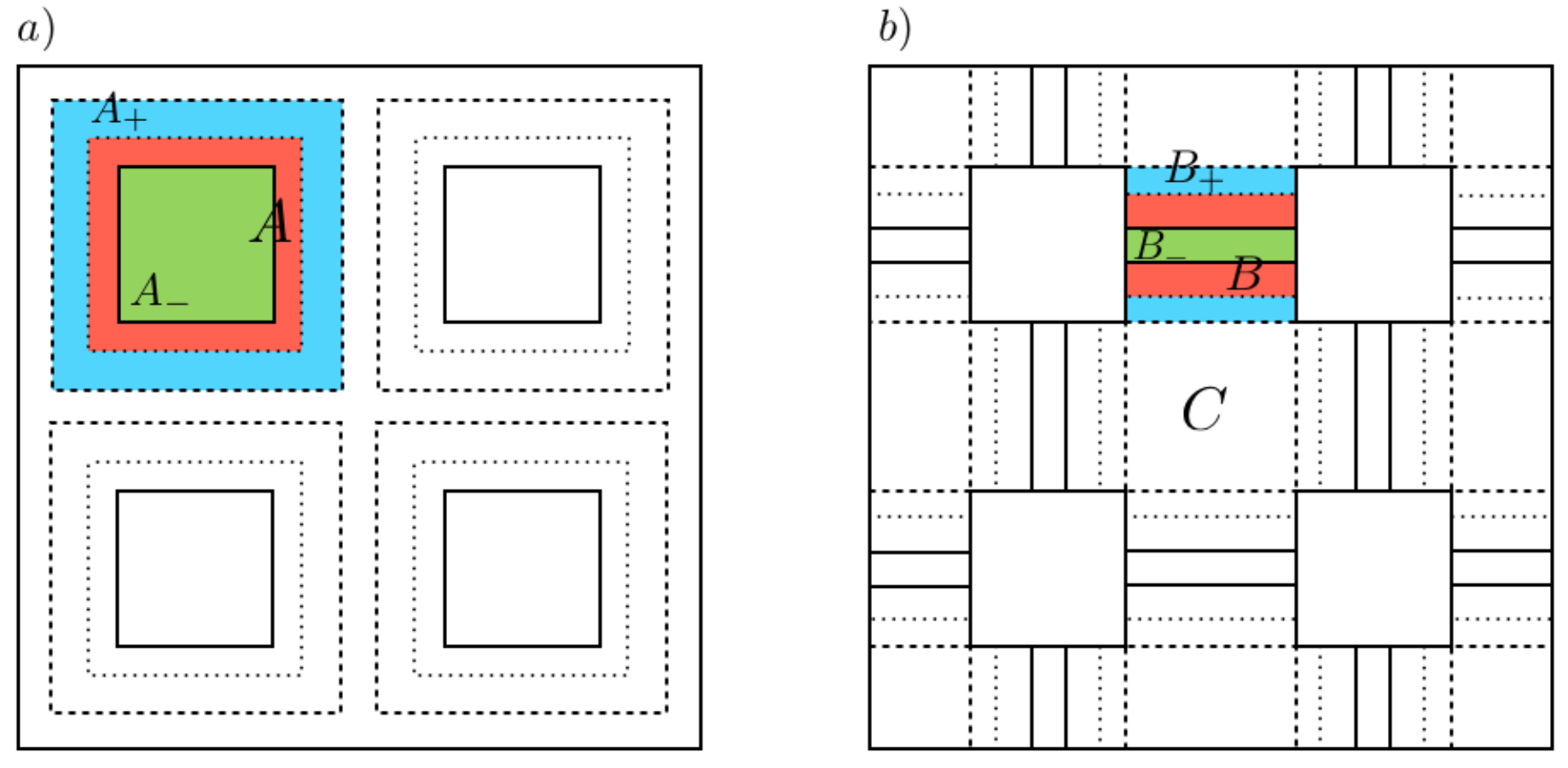}
    \caption{Decomposition of the 2D lattice in the proof of Theorem \ref{thm:main}. In the first step, concentric squares $A_-\subset A\subset A_+$ tile the lattice. $A_-$ is green with a solid line, $A$ is red with a dotted line, and $A_+$ is blue with a dashed line. The boundary of each square is a distance $\ell$ from the boundary of the others other. In the second step b), the inner squares $A_-$ are removed from the lattice, and we consider concentric regions $B_-\subset B\subset B_+$ connecting the inner $A_-$ squares. In the third step, the inner region $B_-$ will be removed, and we are left with a disconnected lattice $\Lambda\setminus(A_-B_-)$. In $D$ dimensions, the procedure needs to be repeated $D+1$ times, removing (0) squares, (1) lines, (2) surfaces, (3) cubes, etc.   }
    \label{fig:ptlat}
\end{figure}

We are now in a position to prove our main result of the paper: an efficient quantum algorithm for preparing Gibbs states of a finite lattice system. There are two natural ways to prepare quantum states on a quantum computer. The first is to prepare a pure state of a system and bath composite, by a local unitary circuit or an adiabatic evolution, and then trace out the bath degrees of freedom. The second method is to prepare the mixed state as the output of a \textit{quantum channel} (completely positive and trace preserving map) by engineered dissipation \cite{disseng}. Ours will be the latter approach.

We will assume that a quantum channel acting on a finite number of neighbouring qubits can be prepared efficiently (with constant time and energy resources) on a quantum computer by engineered dissipation. Furthermore, two quantum channels $T_1$ and $T_2$ acting on non-overlapping sets of qubits can be applied in parallel as $T_1\otimes T_2$, incurring no extra time cost, in very much the same way as two unitaries acting on non-overlapping sets of qubits can be applied in parallel. In analogy to the unitary case, we will say that a channel $T$ is a depth $\kappa$ mixed quantum circuit if it can be written as a product of at most $\kappa$ channels $T=T_\kappa T_{\kappa-1}\cdots T_1$, where each individual channel $\{T_j\}$ can be written as a tensor product of maps acting on non-overlapping sets of neighbouring qubits. We will say that the local channels have \textit{range} $\ell$ if they act non-trivially within a hypercubic box of side length $\ell$.   

We can now state the main theorem.

\begin{thm}\label{thm:main}
Let $H$ be a local bounded Hamiltonian on a $D$ dimensional hypercubic lattice $\Lambda$ and let $\rho$ be its Gibbs state at inverse temperature $\beta$. If $(H, \beta)$ is uniformly $\epsilon(\ell)$-clustering and is uniformly $\delta(\ell)$-Markov, then there is a quantum channel $\bF$ which is a depth $D+1$ mixed quantum circuit $\bF^{(D+1)}$ composed of local channels with range $O(\ell)$ such that
\begin{equation}
\Vert \bF^{(D+1)} \left( \sigma \right)  - \rho  \Vert_1 \leq c D |\Lambda| (\delta(\ell) +\epsilon(\ell/2)+\gamma(\ell/2)),\label{eqn:mainbd}
\end{equation}
for some constant $c$ of order one, and any input state $\sigma$. 
\end{thm}

\proof{The proof is presented for a 2D geometry for illustration purposes, but extends naturally to higher dimensions. There will be two natural length scales in our system $r$ and $\ell$, both independent of $L$, and we assume that $r>9\ell$. We consider a tiling of the lattice (see Fig. \ref{fig:ptlat}a) with $N_A$ concentric squares $A^j_{-}\subset A^j\subset A^j_{+}$, where each $A^j_+$, $j=1,...,N_A$ has side length $r$. $A^j_{-}$ is the inside of $A^j$, a distance $\ell$ inside $A^j$, and $A^j_{+}$ is the outside of $A^j$, a distance $\ell$ outside of $A^j$. Hence there will be $N_A=O((L/r)^2)$ many non-overlapping square $A$-tiles, where $L:= |\Lambda|^{1/2}$ is the lattice side length. Define the quantum channel $\bF^\rho_A=\prod_{j=1}^{N_A}\bF_{A^j}$ with $\bF_{A^j}:=\cR_{A^j_{+}\setminus A^j}^{A^j_{+}}\circ \Tr_{A^j}$, where $\cR_{A^j_{+}\setminus A^j}^{A^j_{+}}$ is the recovery map for the Gibbs state $\rho$ associated with the conditional mutual information $I_\rho(A^j:(A^j_{+})^c|A^j_{+}\setminus A^j)$. Since $\rho$ is $\epsilon(\ell)$-clustering, and the outer boundary of $A^j$ has width $\ell$ we get from Def. \ref{def:markov}:
\be I_\rho(A^j:(A^j_{+})^c|A^j_{+}\setminus A^j)\leq \delta(\ell)\ee
 Hence, from Theorem \ref{thm:FR}, and the union property of recoverable regions,
\be || \bF^\rho_A(\rho)-\rho||_1\leq N_A \delta(\ell).\label{eqn:regionA}\ee
In the next step, we want to compare recovery of the $A$ patches from the original state $\rho$ with recovery of the $A$ patches from modified states $\rho^{A^c_{-}}\propto e^{-\beta H^{A^c_{-}}}$. Then,  
\be||\Tr_{A}(\rho-\rho^{A_{-}^c}\otimes \sigma_{A_-})||_1\leq cN_A|\partial A^{\rm max}_{-}|(\epsilon(\ell/2)+\gamma(\ell/2)) ,\label{eqn:Alocin}\ee
from Theorem \ref{thm:local_indist} and the union property of local indistinguishability, where $|\partial A^{\rm max}_-|$ is the maximal size of any boundary of $A^j_-$, and $\sigma$ is an arbitrary state on $\Lambda$.

Now, recalling that $\bF^\rho_A\circ\Tr_A=\bF^\rho_A$, and combining Eqns. (\ref{eqn:regionA}) and (\ref{eqn:Alocin}), we get
\be || \bF^\rho_A(\rho^{A_{-}^c}\otimes \sigma_{A_-})-\rho||_1\leq cN_A(\delta(\ell) +|\partial A_{-}^{\rm max}|(\epsilon(\ell/2)+\gamma(\ell/2))),\label{eqn:step1}\ee
for some constant $c$. It is now clear that we need the inner and outer boundaries of $A$. The inner boundary provides a shield in order to use local indistinguishability from Theorem \ref{thm:local_indist}. The outer boundary is necessary to allow for the union property of recoverable regions, using the locality of the recovery operations. In short, $\bF_A^\rho=\prod_{j=1}^{N_A} \bF_{A^j}$ is a tensor product of quantum channels acting on regions $A^j_+$ of side length $r$.  Eqn. (\ref{eqn:step1}) tells us that the channel $\bF_A^\rho$ acting on the state $\rho^{A_{-}^c}\otimes \sigma^{A_-}$ recovers the state $\rho$ up to an error $cN_A(\delta(\ell) +|\partial A_{-}^{\rm max}|(\epsilon(\ell/2)+\gamma(\ell/2)))$.

The next step of the proof is to prepare $\rho^{A_{-}^c}$. We will repeat the above step, but with different regions. Define the rectangles $B^j_{-},B^j,B^j_{+}$ as in Fig. 2b), where each $B^j$, $j=1,...,N_B$ connects two squares in $A$, and the inside and outside are defined in the same way as for the $A$ regions with buffer regions of width $\ell$. There will similarly be $N_B=O((L/r)^2)$ many  $B$-tiles.  Define the new quantum channel $\bF^{\rho^{A_{-}^c}}_B=\prod_{j=1}^{N_B}\bF_{B^j}$ with $\bF_{B^j}:=\cR^{B^j_{+}}_{B^j_{+}\setminus B^j}\circ \Tr_{B^j}$, where now $\cR^{B^j_{+}}_{B^j_{+}\setminus B^j}$ recovers the state $\rho^{A_{-}^c}$ from an erasure of $B^j$ according to the conditional mutual information $I_\rho(B^j:(B^j_{+})^c|B^j_{+}\setminus B^j)$. By the same arguments as above, we get 

\be || \bF^{\rho^{A_{-}^c}}_B(\rho^{(A_{-}B_{-})^c}\otimes \sigma_{A_-B_-})-\rho^{A_{-}^c}\otimes \sigma_{A_-}||_1\leq cN_B(\delta(\ell) +|\partial B^{\rm max}_{-}|(\epsilon(\ell/2)+\gamma(\ell/2))),\ee
for some constant $c$. Combining the preparation of regions $A$ and $B$, we get

\be  ||\bF^\rho_A\bF^{\rho^{A_{-}^c}}_B(\rho^{(A_{-}B_{-})^c}\otimes \sigma_{A_-B_-})-\rho\otimes \sigma_{A_-}||_1\leq c(N_A |\partial A^{\rm max}_{-}|+N_B |\partial B^{\rm max}_{-}|)(\delta(\ell) +\epsilon(\ell/2)+\gamma(\ell/2)),\ee
where again $c$ is some constant, and we have loosened the bound only for notational purposes. 
Finally, define the region $C=(A_{-}B_{-})^c$ as shown in Fig. 2b). It is clearly the disconnected union of regions that are at least a distance $\ell$ apart. We can prepare $\rho^C$ from any initial state with the map  $\bF^{\rho^C}_C=\rho^{C}\Tr_C$. Thus, the final sequence of quantum channels reads:
\be ||\bF^\rho_A\bF_B^{\rho^{A_-^c}}\bF^{\rho^C}_C(\sigma)-\rho||_1\leq c(N_A |\partial A^{\rm max}_{-}|+N_B |\partial B^{\rm max}_{-}|)(\delta(\ell) +\epsilon(\ell/2)+\gamma(\ell/2)).\label{eqn:circuitT},\ee
for any input $\sigma$.

Each quantum channel $\bF^\rho_A$, $\bF_B^{\rho^{A_-^c}}$, and $\bF^{\rho^C}_C$ is the product of non-overlapping channels with support in a box of side length $r$, hence $\bF$  forms a depth three mixed quantum circuit of channels. If we take $r>9\ell$, the only relevant length scale in the system is $\ell$. Given that $N_A=O(L^2/\ell^2)$ and $|A^{\rm max}_-|=O(\ell^2)$, and similarly for $B$, we get the scaling pre-factor $O(L^2)=O(|\Lambda)|$ in front of the $(\delta(\ell) +\epsilon(\ell/2)+\gamma(\ell/2))$ error term.
So far our proof has been specific to $D=2$, however it extends naturally to higher dimensions without any modification. The decomposition of the lattice into (0) holes, (1) lines connecting the holes, (2) surfaces connecting the lines, (3) volumes connecting the surfaces, etc. can be extended to any dimension. Only the first and last step of the construction are special in that the last step is simply a projection onto a product state, while the first is special, because it only requires  clustering and the Markov property for contractible regions. Extending the proof to $D$ dimensions yields the bound in Eqn. (\ref{eqn:mainbd}).
\qed}

\subsection{Implications and strengthening of Theorem \ref{thm:main}}\label{sec:implications}

\paragraph{One dimension:}
Classical systems always satisfy the uniform Markov property, and there is no distinction between clustering of contractible and non-contractible regions because of the absence of topological order. It is natural to ask whether or not the assumptions going into the theorems are ever satisfied for quantum systems. 
In one spacial dimension, our theorem holds without any further assumptions. Indeed, a one dimensional system can be covered by  two local overlapping quantum channels. In the proof of Theorem \ref{thm:main}, neither the outmost channel (what we called $\bF^\rho_A$) nor the last channel require  clustering of non-contractible regions, nor in fact uniform clustering. Clustering of the full state $\rho$ is sufficient. Similarly, the  Markov condition is needed only for contractible regions. 

\begin{figure}
\centering
  \includegraphics[scale=0.35]{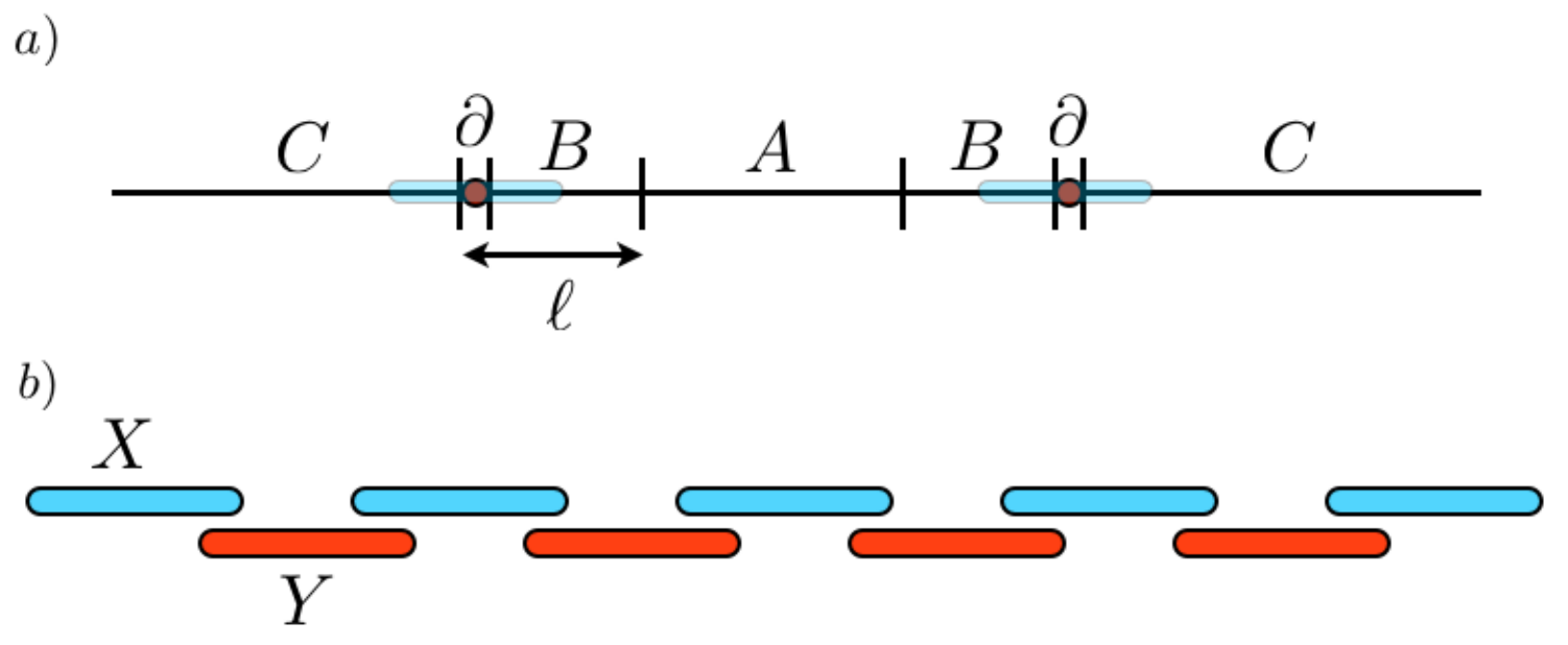}
    \caption{Decomposition of the 1D lattice. Details described in the main text.}
    \label{fig1D}
\end{figure}

Indeed, consider the one dimensional line in Fig. \ref{fig1D}, where the lattice $\Lambda=AB\partial C$, $\partial_L$ and $\partial_R$ are a single site constituting the boundary between $B$ and $C$. As in Theorem \ref{thm:local_indist}, we get that $||\Tr_{BC}(\rho^X)-\Tr_{B}(\rho^{AB})||_1\leq c (\epsilon(\ell/2)+\gamma(\ell/2))$ for some constant $c$ if $\Cov_{\rho^{ABC}}(f,g)\leq\epsilon(\ell)$ for any observables $f$ with support in $A$ and $g$ with support in $C$, since the boundary $|\partial |$ is constant. 
It is not difficult to show that $\Cov_{\rho^\Lambda}(f,g)\leq\epsilon(\ell)$ implies $\Cov_{\rho^{ABC}}(f,g)\leq\epsilon(\ell)$: 
\bea \Cov_{\rho^{\Lambda}}(f,g)&=&|\tr{\rho^{\Lambda} f g}-\tr{\rho^{\Lambda}f}\tr{\rho^{\Lambda}g}|\nonumber\\
&\leq& |\tr{\eta_{\ell/2}^{\partial}\rho^{ABC} \eta_{\ell/2}^{\partial,\dag} f g}-\tr{\eta_{\ell/2}^{\partial}\rho^{ABC} \eta_{\ell/2}^{\partial,\dag} f}\tr{\eta_{\ell/2}^{\partial}\rho^{ABC} \eta_{\ell/2}^{\partial,\dag} g}|+5\gamma(\ell/2)||f||~||g||\nonumber\\
&=& |\tr{\rho^{ABC} f \tilde{g}^1_{\ell/2}}-\tr{\eta_{\ell/2}^{\partial,\dag}\eta_{\ell/2}^{\partial}\rho^{ABC}  f}\tr{\rho^{ABC} \tilde{g}^1_{\ell/2}}|+5\gamma(\ell/2)||f||~||g||\nonumber\\
&\leq& (2\epsilon(\ell/2)+5\gamma(\ell/2))||f||~||g||,\eea
where $||\eta_{\ell/2}^{\partial}\rho^{ABC} \eta_{\ell/2}^{\partial,\dag}-\rho^\Lambda||_1\leq\gamma(\ell/2)$ by belief propagation, and the operators $\eta_{\ell/2}^{\partial}$ are localized a the boundary $\partial$. Hence they are local and bounded operators. We have defined the operators $\tilde{g}_{\ell/s}= \eta_{\ell/2}^{\partial,\dag}g \eta_{\ell/2}^{\partial}$ that have support on a ball of radius $\ell/2$ around the boundary $\partial$. 

We now can apply the depth two quantum channel to any input state $\sigma$ to get
\be ||\bF^\rho_X\bF^{\rho_Y}(\sigma)-\rho^\Lambda||_1\leq c N_X(\epsilon(\ell/2)+\gamma(\ell/2)+\delta(\ell)),\ee
using the same construction as in the proof of Theorem \ref{thm:main} (see Fig. \ref{fig1D}). 
The 1D proof does not require a uniform clustering or the uniform Markov property, but it does require them to hold for the full state of the lattice.  

It has been known for a while that one dimensional Gibbs states are clustering \cite{araki1969}. Recently, one of the authors has shown that Gibbs states of one dimensional lattice systems always satisfy the Markov condition \cite{KatoBrandao16}. Thus, quantum Gibbs states of one dimensional lattice systems can be efficiently prepared on a quantum computer at any finite temperature, without any further assumptions.

\paragraph{Finite temperature topological order:}

Our results provide some insight into the problem of self-correcting quantum memories based on commuting projector codes (local commuting Hamiltonians) \cite{brown2014}. A self-correcting quantum memory is expected to be associated with robust topological order at finite temperature \cite{hastings2011}. It has been argued that this should be related to long lived (metastable) states with topological order, or in other words, that there exists a protected subspace into which quantum information can be reliably stored. Most proposals for self-correcting quantum memories have been stabilizer models \cite{dennis2002,haah2011,michnicki,brell2016}, although other proposals do exist; see Ref. \cite{brown2014} for a recent review. For those models, the uniform Markov condition always holds, as the Hamiltonian is commuting.  

Therefore, Theorem \ref{thm:main} holds whenever uniform clustering holds. It is known that the $4D$ toric code at finite temperature is a self-correcting quantum memory \cite{alicki2010}, and it is not clustering\footnote{MB. Hastings, private communication}. It is an interesting open question to understand whether there exist quantum models satisfying uniform clustering for contractible regions, but not for non-contractible regions, in the same way as the Toric code ground state does.

\paragraph{Ground States:}

The strategy underlying the proof of Theorem \ref{thm:main} extends to more general classes of states than Gibbs states of local Hamiltonians. The basic ingredients are: (i) the existence of local recovery maps from the Markov condition on non-contractible regions, and (ii) local indistinguishability on non-contractible regions. In both of these ingredients, there is a tacit property that is invoked; that the class of states can be meaningfully defined on arbitrary deformations of the lattice. In our case, the deformations consist of removing sites. We can meaningfully speak of all of these properties for ground states of certain gapped Hamiltonians. 

Indeed, let $H$ be a local bounded frustration-free Hamiltonian on $\Lambda$ and define the restricted Hamiltonian $H^A$ as above for $A\subset\Lambda$, and let $\cG^A$ be its set of ground states. Because of frustration freedom, $\cG^\Lambda\subset\cG^A$. $H$ is said to satisfy local indistinguishability if
\be ||\Tr_{BC}(\psi)-\Tr_B(\phi^{AB})||_1\leq c e^{-\dist{A,C}/\xi},\ee
for any $\psi\in\cG^\Lambda$, $\phi^{AB}\in\cG^{AB}$ and $c, \xi$ are constants. 

The Markov condition in this case corresponds to the assumption that the topological entanglement entropy (TEE) is zero. Assuming TEE is zero and local indistinguishability, we can now prove  Theorem \ref{thm:main} in exactly the same way as for Gibbs states (we explicitly have to use local indistinguishability, since Theorem \ref{thm:local_indist} generally does not extend to pure states\footnote{It is worth noting however that for injective PEPS with a uniformly gapped parent Hamiltonian, Theorem \ref{thm:local_indist} can be shown in much the same way as in Ref. \cite{schwarz2016}.}). One subtlety to keep in mind in the pure state case is that if the ground subspace is degenerate, then the state preparation will depend on the input state to the circuit. This subtlety does not occur in the Gibbs state setting, as the Gibbs state is always unique for finite dimensional systems.

The situation becomes particularly interesting when we consider the preparation of pure topologically ordered states based on commuting projector codes, such as the Toric \cite{kitaev2003} or Color \cite{bombin2006} codes. For such systems, we know that the Markov property is satisfied on contractable regions, and that the ground states are clustering \cite{BPT}. The error functions for these systems are step functions; i.e. $\epsilon(\ell)=\delta(\ell)=0$ beyond a certain length scale.  However, because the topological entanglement entropy is non-zero, the Markov property does not hold on non-contractable regions, and nor does clustering. Hence Theorem \ref{thm:main} does not hold, which is expected since topologically ordered states cannot be prepared by a finite depth circuit by definition. Conversely, for commuting projector codes, if the topological entanglement entropy is zero, then the Markov property on non-contractible regions and uniform clustering do hold, and hence we can construct a finite depth circuit of quantum channels that  prepares the states. A similar unpublished result was discussed in a talk by A. Kitaev \cite{KitaevTalk}. Injective PEPS satisfying uniform clustering are a class of states for which the assumptions of our theorem are expected to hold.

\section{Outlook}
In this paper, we have shown two theorems on the efficiency of (a) estimating local observables of and (b) preparing the Gibbs state of a local Hamiltonian assuming that the state has exponential decay of correlations, and that it satisfies the approximate Markov condition uniformly on non-contractible regions. The efficiency of our algorithm is essentially optimal since it is impossible in general to have a strictly constant depth circuit with arbitrarily small error if there are non-zero (though possibly exponentially small) correlations at distances of order $O(|\Lambda|)$. We provide the first Gibbs state preparation algorithm that has convergence guarantees which are checkable based only on static properties of the system. These guarantees, uniform clustering and the Markov condition, are physically motivated. The proof strategy does not depend too decisively on the structure of Gibbs states, and we expect that the methods developed here will be useful for the preparation of other classes of pure and mixed states. 

A number of open problems associated with the present work  remain to be solved. Perhaps the most pressing open problem is to know under what conditions the Markov property holds for Gibbs states of non-commuting Hamiltonians in dimensions grater than one. It is the belief of the authors that if a Gibbs state is clustering, then it satisfies the Markov condition, and if it is uniformly clustering then it satisfies the uniform Markov condition. We have unfortunately not been able to prove this claim. A second pressing question, and one of perhaps greater physical relevance, is to understand whether and under what conditions there can be a separation between clustering and/or the Markov property for contractable and non-contractible regions. In the case of ground states, there is a separation between these two regimes, for which topological order is responsible. Understanding whether such a separation is connected to topology in the finite temperature setting is an important open problem. 

Another exciting direction to explore is whether it is possible to replace the finite sequence of channels with $\log(N)$ depth support with a $\log(N)$ depth sequence of channels with constant support. The analysis of a local semigroup reflecting a non-ommuting Gibbs sampler \cite{GibbsSampling} would be a possible approach. The criterion for convergence would then be the mixing time of the semigroup. The advantage of such an approach would be twofold. First, it would show that the local circuit preparing the state does not need to be properly time ordered, but rather can be applied asynchronously. Second, and more importantly, this would give us a new avenue for proving gaps of semigroups, in the spirit of the classical analysis of Glauber dynamics \cite{ziggy,martinelli}. This in turn could potentially be translated to the ground state setting to provide new tools for proving gaps of frustration-free Hamiltonians. 

\noindent \textit{Note added:} After the completion of this work we learned about the papers by Swingle and McGreevy \cite{swingle2015,swingle2016}, which also studies the problem of preparing quantum states efficiently using similar techniques from ``patching". In particular, in Ref. \cite{swingle2016} they explore the implications of a vanishing conditional mutual information for the task.

\paragraph{Acknowledgements:}
We thank Angelo Lucia and David Perez-Garcia for helpful discussions. We thank Isaac Kim for pointing out the unpublished result of Kitaev to us. 
MJK was supported by the Carlsberg fund and the Villum fund.

\bibliographystyle{unsrt}


\end{document}